\documentclass{aa}
\usepackage{natbib}
\usepackage{graphicx}
\usepackage[english]{babel}
\usepackage{txfonts}
\usepackage{hyperref}
\usepackage{tabularx,booktabs}
\usepackage{multirow}
\usepackage{adjustbox}

\bibpunct{(}{)}{;}{a}{}{,}

\newcommand{\Msun}{$M_{\sun}$}

\begin{document}

\title
{Accretion and outflow activity on the late phases of pre-main-sequence evolution. The case of RZ Piscium.}

\author
{I.S.\,Potravnov\inst{1} \and D.E\, Mkrtichian\inst{2} \and V.P.\, Grinin \inst{1,3} \and
I.V.\,Ilyin\inst{4}
\and D.N.\, Shakhovskoy\inst{5}} \institute
{Pulkovo Astronomical Observatory, Russian Academy of Sciences, 196140, Pulkovo, St.\,Petersburg, Russia\\
e-mail: ilya.astro@gmail.com \and National Astronomical Research Institute of Thailand,
Chiang Mai 50200, Thailand \and Saint-Petersburg State University, Universitetski pr. 28, 198504  St.\,Petersburg, Russia
\and Leibniz-Institut f\"{u}r Astrophysik Potsdam (AIP), An der Sternwarte 16, 14882 Potsdam,
Germany  \and Crimean Astrophysical Observatory, P. Nauchny, 298409 Republic of Crimea
\\
}

\date{Received 13 May 2016/accepted 02 December 2016}

\titlerunning{ }

\authorrunning{Potravnov et al.}

\abstract {RZ Psc is an isolated high-latitude post-T Tauri star that demonstrates a UX Ori-type photometric activity.
The star shows very weak spectroscopic signatures of accretion, but at the same time possesses the unusual footprints of the wind in \ion{Na}{I} D lines. 
In the present work we investigate new spectroscopic observations of RZ Psc obtained in 2014 during two observation runs.
We found variable blueshifted absorption components (BACs) in lines of the other alcali metals, \ion{K}{I} 7699 \AA\ and \ion{Ca}{II} IR triplet.
We also confirmed the presence of a weak emission component in the H$\alpha$ line, which allowed us to estimate the 
mass accretion rate on the star as $\dot{M}\leq$ $7 \cdot 10^{-12}$\Msun yr$^{-1}$. 
We could not reveal any clear periodicity in the appearance of BACs in sodium lines. 
Nevertheless, the exact coincidence of the structure and velocities of the \ion{Na}{I} D absorptions observed with the interval of about one year suggests 
that such a periodicity should exist.}

\maketitle

\section{Introduction}

It is well known from observations that many young stars are surrounded by the circumstellar disks. During their evolution disks dissipate partly 
owing to gas accretion onto the star, partly in the course of the photoevaporation and planets formation (see reviews by \citealt{Williams_Cieza2011,Alexander2014}). 
In the case of low-mass T Tauri stars (TTS), these processes are well studied both from observational and theoretical points of view (see \citet{Bouvier2007} and 
references therein). The interaction between the stellar magnetosphere and circumstellar disk plays a crucial role in the accretion process. The bulk of the accreting gas 
infalls onto the star and only a minor part ($\sim$ 10\%) leaves the stellar vicinity via magnetospheric conical and X-winds (see review by \citealt{Ferreira2013} and references therein). 
The observational signatures of this process are well known: strong emission in the Balmer and some metallic lines, and their typical P Cyg profiles. 

Through the evolution of protoplanetary disks and the dissipation of gaseous matter in their inner parts, the accretion, 
and hence outflow, gradually decays. The characteristic time 
of this decay is several millions years (Myr). On the timescale of about 10 Myr accretion becomes vanishingly small (less than $10^{-11}$\Msun yr$^{-1}$) 
\citep{Fedele2010}. However, such a short timescale probably arises as a result of selection effects. 
In some young objects the time of disk dissipation is prolonged above this limit \citep{Pfalzner2014}. Nevertheless, the simple rule is that 
there are no signs of outflow if there is no active accretion in the system. Until recently, this rule was applicable without any exceptions. 
A few years ago we found an interesting exception: the unusual variable star RZ Psc.

RZ Psc is classified as a UX Ori-type star on the basis of photo-polarimetric observations: most of the time the star is in a 
bright state (V $\sim$ 11$^m$.5), but it occasionally falls into a deep (up to $\Delta$V $\sim$ 2$^m$) and very short (up to 2$^d$) Algol-like minima
\citep{Zajtseva1985}. The "blueing effect" and the anticorrelation between the stellar brightness and linear polarization degree are also observed in the deep minima
\citep{Kiselev1991, Shakhovskoy2003}.

RZ Psc is a relatively young star: we found the prominent \ion{Li}{I} $\lambda$6708 line in its spectrum \citep{Grinin2010} and our age estimation,
based on a\ kinematical approach, is $25\pm 5$ Myr \citep{Potravnov2013kin}. 
RZ Psc shows strong mid-IR excess ($\lambda \gtrsim $ 3 $\mu$m) and absence of an excess of emission at shorter wavelengths \citep{deWit2013}.  
This means that the circumstellar disk of RZ\ Psc has an inner gap with the radius of several tenths of AU. 
The blackbody radiation with a temperature of about 500 K fits the observed excess well, and the fractional 
luminosity of the dust in this case is $\sim 8\%$. The large amount of warm dust was interpreted by \citet{deWit2013} in the framework 
of the debris disk model as a result of possible recent collisional events in the planetesimal belt.

Commonly, the debris disks are considered gas-poor disks (see reviews by \citealt{Matthews2014, Wyatt2008} and references therein).
Only small amounts of cold gas are observed in \element[][][][]{CO} lines or via weak spectroscopic signatures
of evaporating comets as in the case of 23 Myr old \citep{Mamajek2014} $\beta$ Pic and few other A-type stars with debris disks \citep{Welsh2013}.

In contrast, RZ Psc shows perceptible circumstellar activity.
At a first glance, the spectrum of the star looks like the spectrum of a late-type main-sequence (MS) star.
The spectrum does not show any emission above the continuum, and there is only very weak emission (EW $\sim$ 0.5 \AA) in the core of the H$\alpha$ line. 
At the same time the star possesses the prominent and variable signatures of the matter outflow in \ion{ Na}{I} D resonance lines \citep{Potravnov2013gas}.
Such outflow activity is not typical for stars with debris disks, but this activity can be expected in the late evolutionary phases of stars with primordial disks.

We suppose that the unusual spectral variability of RZ Psc is caused by the interaction between an inclined stellar magnetosphere 
and remnants of the accreting gas in the magnetic propeller regime \citep{Grinin2015}.
In this instance one can expect the modulation of variable details in \ion{Na}{I} D lines with a period that is close 
to the rotational period of the star. The identification of the recurrent 
details in \ion{Na}{I} D lines, on the basis of the new spectroscopic observations of RZ Psc, is the aim of the present work.

\section{Stellar parameters}

In this section we describe and update some of the basic stellar parameters.
The parameters of RZ Psc atmosphere were derived using high-resolution spectroscopy obtained with the Nordic Optical Telescope \citep{Potravnov2014}. 
These data are summarized, along with other data, in Table ~\ref{table2}.

\begin{table}
\caption{Parameters of RZ Psc}
\label{table2}
\centering
\begin{tabular}{c c c }
\hline\hline
\addlinespace
Parameter & Value & Reference  \\
\hline
\\
   $T_{\rm eff}$ & 5350$\pm$150K & 1 \\
   $\lg g$ & 4.2$\pm$0.2& 1\\
   $[M/H]$ & -0.3$\pm$0.05& 1 \\
   $V\sin$ \emph{i} & 12.0$\pm$0.5 km/s& 1\\
   $R_{\star}$/$R_{\sun}$ & 0.9 & 2\\
   $M_{\star}$/$M_{\sun}$ & 1.0 & 2\\
   $L_{\star}$/$L_{\sun}$ & 0.7 & 2\\
   EW(\ion{Li}{I}) & 0.202\AA & 3\\
   EW(H$\alpha$ emission) & 0.5\AA & 2,4\\
   $\dot{M}$& $\leq$ $7 \cdot 10^{-12}$\Msun yr$^{-1}$ & 2\\
   Galactic latitude & -35$^{\circ}$ & 3\\
   $RV$& -1.2 $\pm$0.33 km/s & 5 \\
   
\hline
\end{tabular}
\tablefoot{References: 1- \citet{Potravnov2014}; 2- this work, 3 - \citet{Grinin2010}; 4 - \citet{Grinin2015}; 5 - \citet{Potravnov2014b} }
\end{table}

\begin{figure*}
\begin{centering}
\includegraphics[width=0.8\linewidth, angle =0]{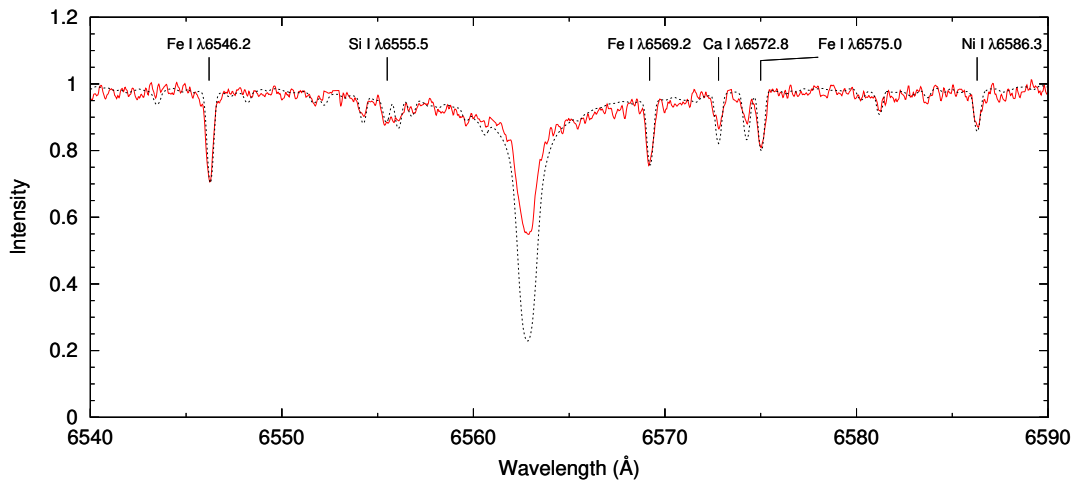}
\caption{\label{2} Comparison of the region around H$\alpha$ line in RZ Psc spectrum 
(solid light gray line or red in the electronic version of the paper) with the same line in
the $\sigma$ Dra spectrum (dotted line). 
The spectrum of RZ Psc was obtained on 2013 November 21 with the FIES spectrograph on the Nordic Optical Telescope (R $\sim$ 46000). The spectrum
of $\sigma$ Dra (R $\sim$ 42000) was retrieved from the ELODIE archive at Observatoire de Haute-Provence \citep{Moultaka2004}.}
\end{centering}
\end{figure*}

Figure~\ref{2} compares the RZ Psc spectrum in the vicinity of the H$\alpha$ line with that of the K0 V standard $\sigma$ Dra \citep{Keenan1989}.
We used $\sigma$ Dra since its atmospheric parameters, including its metallicity \citep{Mishenina2013}, were close to those determined for RZ Psc. 
Taking into account the slow
rotation of $\sigma$ Dra, which is $V\sin i$ = 1.4 km s$^{-1}$ \citep{Marsden2014}, we convolved its spectrum with the corresponding rotational kernel to 
adjust the rotational broadening of the lines observed in the RZ Psc spectrum ($V\sin i$ = 12 km s$^{-1}$). One can see the good coincidence of the photospheric
 \ion{Fe}{I},  \ion{Si}{I,} and  \ion{Ni}{I}  lines in the both spectra which, in general, extends over the whole available wavelength range
 (except the lines of the alkali metals, see the next section).
On the other hand, the H$\alpha$ line in the RZ Psc spectrum looks different from the standard spectrum. While the outer parts of the line wings coincide 
well with the standard ones, the line core is shallower than that in the standard spectrum. This can be result of the chromospheric activity, weak accretion, or
a combination of both of these effects (see Section 5). The spectrum of the star indicates that RZ Psc has already passed the stage of an actively accreting classical TTS and now 
appears spectroscopically to be a very weak-line TTS or post-TTS. 

In our previous paper we estimated the age of RZ Psc ($25\pm 5$ Myr) using the calculations of its space motion under the assumption 
of the Galaxy plane as the starting point. 
However, one should take this estimate with some caution because the exact birthplace of RZ Psc  in the Galaxy is presently unknown and the 
precision of the calculations is limited by the uncertainties of the input data, in particular, the distance.

\begin{figure}
\includegraphics[width=\linewidth, angle =0]{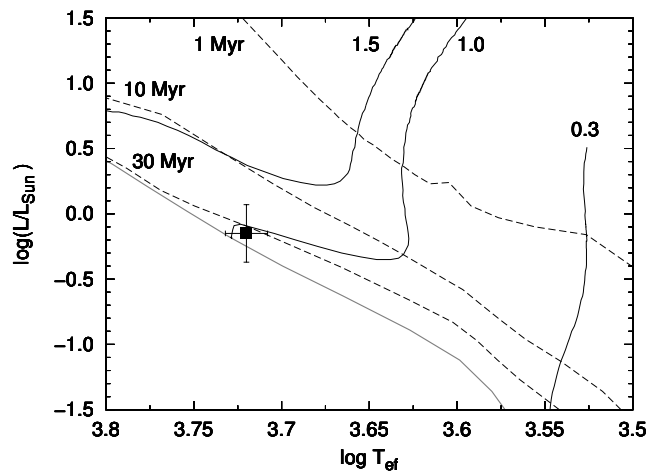}
\caption{\label{1} RZ Psc (filled square) on the HR diagram. The errors indicate the precision of the determination of the $T_{\rm eff}$ and $\lg g$.
The light gray solid line is the ZAMS. Thin solid lines are the evolutionary tracks from \citet{Siess2000} models, these tracks are labeled with the 
corresponding values of the stellar masses. The dashed lines are the isochrones for the ages 1, 10, and 30 Myrs.}
\end{figure}

In the present work we adopt the distance value $D\sim$160 pc
from \citet{Pickles2010}. 
The extinction in the direction on RZ Psc is $A_{v}= 0.3 \pm 0.05$ (G. Gontcharov, private communication). 
Hence, the luminosity of RZ Psc is about of $0.7 L_{\sun}$. With this luminosity and temperature $T_{\rm eff}$=5350K 
the evolutionary models of \citet{Siess2000} provide the value $\sim 0.9 R_{\sun}$ for the stellar radius and corresponding mass $\sim 1.0 M_{\sun}$.
The position of the star on the HR diagram is shown in the Figure ~\ref{1}. RZ Psc lies at the end of its pre-main-sequence (PMS) track, however, it is still
located above the MS.


Taking the stellar radius $R_{\star}= 0.9 R_{\sun}$ and the projected rotation velocity 
$V\sin$ \emph{i} = 12.0 km s$^{-1}$  into account, we can estimate the rotational period of the star. Since RZ Psc
belongs to UX Ori-type stars, the orientation of its circumstellar disk is close to edge-on. We therefore assume that the inclination angle is $i = 70^{\circ}$,
as in UX Ori itself \citep{Kreplin2016}. 
In this case the rotational period is about P $\sim3.6^{d}$.

\section{Observational data}

The new spectra of RZ Psc were obtained during the two observational runs. The first was carried out by I. Potravnov 
and D. Shakhovskoy in September 2014 at the 2.6 m Shajn telescope of Crimean Astrophysical Observatory with the ESPL echelle spectrograph. 
The ESPL was equipped with a Andor iKon-L 936  CCD detector of $2048 \times 2048$ pixels (pixel size 13.5 $\mu$m). A $4 \times 4$ binning was used. 
A slit width was 2.0 arcsec projected on the sky, which corresponds to a resolving power of $R \sim 22000$ or about 13.5 km s$^{-1}$. 
The exposure covered the spectral region from about 4400 to 7800 \AA~ in 48 orders but with substantial interorder gaps across the whole
wavelength range. These gaps are the result of the fact that the current CCD is smaller than detector for which spectrograph was designed. The typical 
S/N ratio for this observational material was about 50-60 per resolution element. Because of the relatively low S/N ratio, interorder gaps, and 
the impossibility of constructing an accurate global dispersion curve for further cross-correlation radial velocities measurements, we only
analyzed orders that contain \ion{an Na}{I} D doublet.\\ 

The second portion of the observations have been obtained by D. Mkrtichian in December 2014 at the 2.4 m telescope of Thai National Observatory 
with MRES echelle spectrograph. The covered region was from 4400 to 8800 \AA\ in 41 orders with the good overlapping.
The measurements of several unsaturated telluric lines in the region around 5900 \AA\ provide a mean value of FWHM = 0.35~\AA,\ which corresponds to 
resolving power $R \sim 17000$ or about 18 km s$^{-1}$ on the velocity scale. In the same region, S/N was about 95 per resolution element.
A full set of calibration frames was obtained each night. Also the spectrum of the hot fast 
rotating A2V star $\theta$ And was observed for the further telluric correction of science data.

The spectra from both sets were processed in the same manner using the standard IRAF reduction tools. The telluric correction of science data 
was performed using the \textit{telluric} task in IRAF.
Each spectrum was corrected for the instrumental shift of zero point velocity. The values of correction were obtained by measuring the position of the
atmospheric lines.\\

It is important to note that the ESPL exposures have a great advantage in their extension over six successive nights, which covered the
whole suspected rotational period of the star.
The total duration of the MRES observations was 15 days but the spectra obtained were sparsely distributed over this interval. The complete observational 
log is presented in the Table ~\ref{table1}. Although the MRES observations have worse time coverage, the quality of these spectra allows us to measure
the radial velocity of RZ Psc and investigate its spectral variability in the wide wavelength range. The variable lines of alkali metals, \ion{Na}, \ion{K},
\ion{Ca}, as well as the H$\alpha$ line, were simultaneously covered by a single exposure. In terms of understanding the unusual variability in the 
\ion{Na}{I} D lines these data also play an important role in the recognition the typical patterns of the additional components 
and in increasing the statistics of their velocities distribution.

The quasi-simultaneous photometric observations were carried out by D.N.
Shakhovskoy and S.P. Belan using the 1.25 m AZT-11 telescope of the Crimean Astrophysical Observatory. These observations have shown that RZ Psc was in a bright
state during the time of spectral observations, i.e., V $\sim$ 11$^m$.5 .

\begin{table}[h]
\caption{Observational log and radial velocities of RZ Psc.}
\label{table1}
\begin{tabularx}{\linewidth}{l c c c}
\hline\hline
\addlinespace
 Date  & JD  2450000+ & Instrument & RV [km s$^{-1}$]\\
\hline
\\
   2014 Sept 13 & 6914.42 & ESPL & ---\\
   2014 Sept 14 & 6915.43 & ESPL & ---\\
   2014 Sept 15 & 6916.43 & ESPL & ---\\
   2014 Sept 16 & 6917.41 & ESPL & ---\\
   2014 Sept 17 & 6918.43 & ESPL & ---\\
   2014 Sept 18 & 6919.45 & ESPL & ---\\
\addlinespace
\hline
\addlinespace
   2014 Dec 13  & 7005.02 & MRES & -2.22 $\pm$ 0.53 \\
   2014 Dec 15  & 7007.12 & MRES & -1.40 $\pm$ 0.44 \\
   2014 Dec 19  & 7011.12 & MRES & -1.96 $\pm$ 0.53 \\
   2014 Dec 24  & 7016.06 & MRES & -1.54 $\pm$ 0.60 \\
   2014 Dec 25  & 7017.05 & MRES & -1.93 $\pm$ 0.45 \\
   2014 Dec 27  & 7019.07 & MRES & -1.22 $\pm$ 0.60 \\
\\
  
\hline
\end{tabularx}
\end{table}

\subsection{Radial velocities}

Heliocentric radial velocity (RV) of RZ Psc were measured on MRES data using the cross-correlation technique realized in the IRAF \textit{fxcor} task.
The linearized observed spectrum was correlated with the synthetic spectrum in the wavelength range 5000-6500 \AA\ , with the exception of variable \ion{Na}{I} D doublet.
The position of CCF peak was obtaned by fitting to a 
Gaussian profile. These measurements are presented in Table ~\ref{table1}. The typical error, as can be seen from Table ~\ref{table1}, was about $\pm$0.5 km s$^{-1}$. 
The mean value of RZ Psc radial velocity from these spectra was --1.7 km s$^{-1}$. Previously we obtained two very close values of RV ( about --1.2 km s$^{-1}$ ) 
using high-resolution FIES spectra separated by an interval of about three months \citep{Potravnov2014b}. Actually these two values are within the error
limits and  the difference is possibly the result of the offset MRES zero point with respect to the FIES system. Nevertheless, the situation with
RZ Psc RV remains unclear because in the paper cited above we reported about RV fluctuations up to several km s$^{-1}$ observed on the 
different instruments. Subsequent homogeneous observations will possibly clarify the reality of these fluctuations.

Since we investigate the circumstellar gas motions with respect to the star, hereafter we consider the velocities of the spectral features in the rest frame of the star using the RV values obtained above.

\section{Results}

\subsection{Previous spectroscopy of RZ Psc}

The spectral variability of RZ Psc and its general behavior has been found in the moderate resolution spectra from Terskol Observatory  \citep{Potravnov2013gas}. 
The photospheric \ion{Na}{I} D lines were usually flanked by the variable blueshifted absorption components (BACs).  These components completely disappeared some nights. 
It is important to stress that only blueshifted components have been observed; there were no signs of absorptions displaced 
to the red or any signatures of emission in sodium lines.
Valuable information about the fine structure of these BACs was obtained from the two FIES (Nordic Optical Telescope) exposures. 
These observations were partly discussed in \citet{Potravnov2013gas} and \citet{Grinin2015}. 
It was clear from these high-resolution ($R \sim 46000$) spectra that the BACs that were practically unresolved in the Terskol shots  consisted, in reality, 
of one or two narrow discrete components. The velocities of these components varied along with their intensities. 
This variability clearly indicates the circumstellar origin of these discrete components.
After deconvolution with the instrumental profile (6.5 km s$^{-1}$) the FWHM of the narrowest
component was about 12.7 km s$^{-1}$. This value indicates
the very small velocity dispersion in the stream where this absorption component was formed. The ratio of the internal velocity of the stream to its RV was about 1/10.

The FIES spectra also showed the existence of a very weak
variable emission component in the center of the H$\alpha$ line \citep{Potravnov2013gas,Potravnov2014}. The
residual emission, which was obtained after synthetic profile substraction, had EW $\sim$ 0.5 \AA\  and consisted of a narrow peak centered on zero velocity and weak emission wings that extended up to 
$\pm$ 200 km s$^{-1}$ \\

\subsection{\ion{Na}{I} D lines}

\begin{figure}
\center
\includegraphics[width=\linewidth, angle =0]{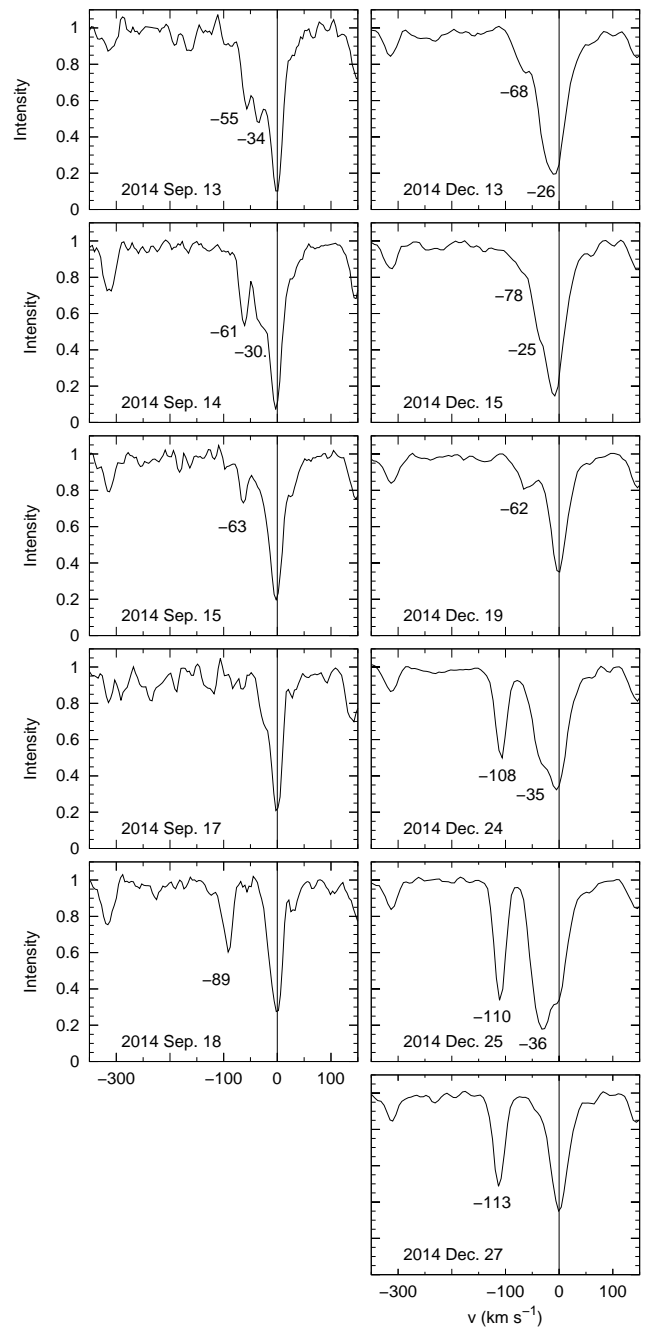}
\caption{\label{3} \ion{Na}{I}$ \lambda$5889 lines in the RZ Psc spectra observed with ESPL (left column) and MRES (right column).
The velocities of the variable absorption components are labeled in km s$^{-1}$.
The weak absorptions near the red edge of $\lambda$5889 line in the ESPL spectra are the residual atmospheric contamination after imperfect telluric correction}
\end{figure}

Figure ~\ref{3} shows the successive evolution of the \ion{Na}{I} $\lambda$5889 profile in the ESPL and MRES  
spectra. We describe these spectra below starting from the earliest ESPL data. 
The first spectrum (of September 13) clearly
demonstrated two separate BACs with velocities of about $-55$ and $-34$ km s$^{-1}$. The velocity of the first component increased up to $-61$ km s$^{-1}$ 
on the following night. The second appeared closer to the photospheric profile. The tenative measurement of its velocity gave a value of about $-30$ km s$^{-1}$. 

In the following spectrum obtained on September 15, the $-30$ km s$^{-1}$ component 
completely disappeared, while the second component stayed near the same position as on the previous night. However the central intensity and width of this
absorption were somewhat reduced. The spectrum obtained on September 16 is not shown in Fig.~\ref{3} because of its poor quality. However, as on
the following plot (September 17), we lacked additional components. Only the slight asymmetry of the blue wing of the photospheric profiles can be noticed. 
The last spectrum observed on September 18 demonstrated the sudden appearance of strong BAC at a velocity of about $-89$ km s$^{-1}$.\\

\begin{figure*}
\begin{centering}
\includegraphics[width=\linewidth, angle =0]{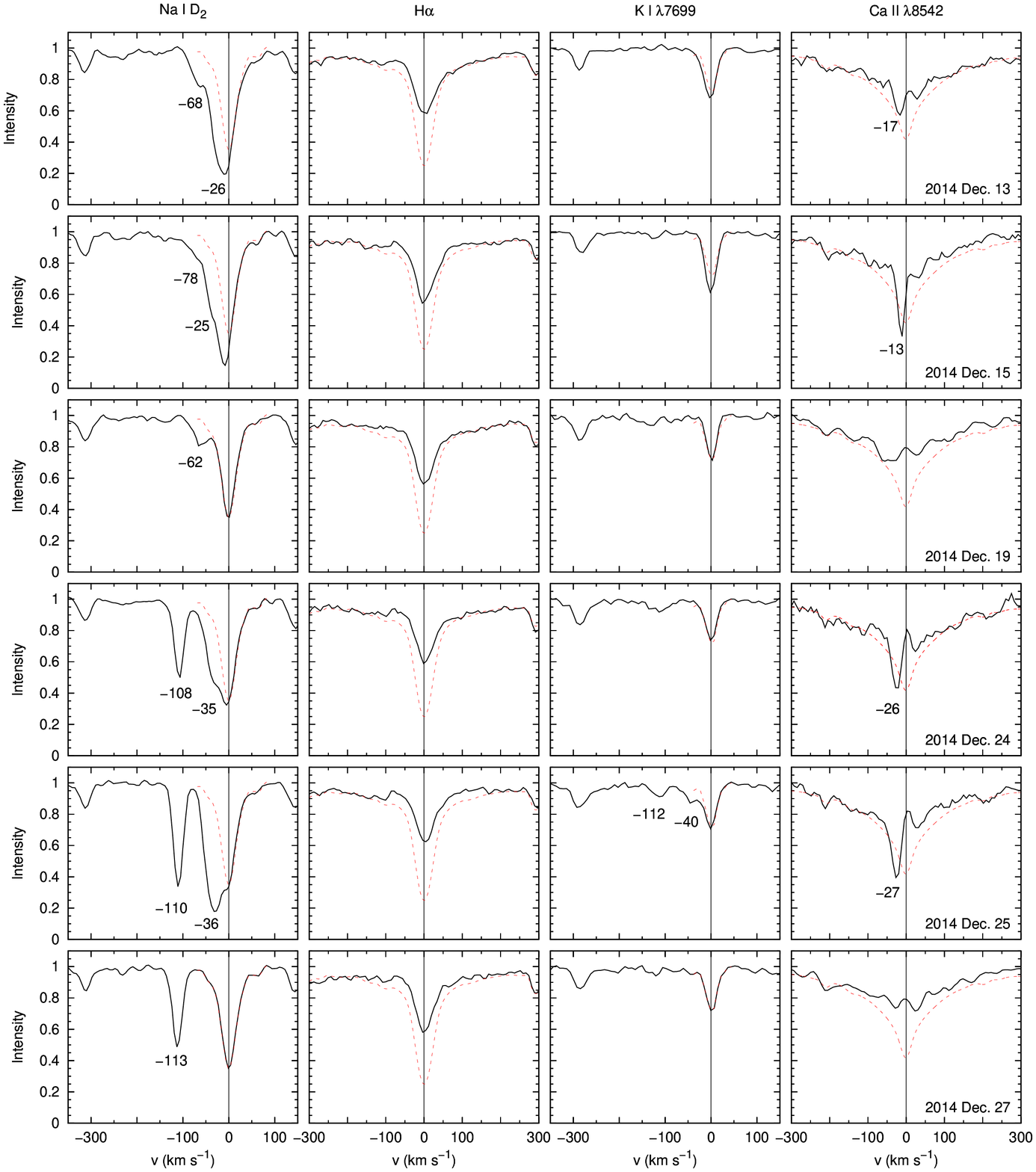}
\caption{\label{5} Profiles of the variable lines observed in the RZ Psc spectra during the MRES run. The velocities in km s$^{-1}$ of the circumstellar
absorptions are labeled. The reference photospheric profiles of the \ion{Na}{I} $\lambda$5889 and \ion{K}{I} $\lambda$7699 
are plotted with a dashed (red in the electronic version) line (see the text for details). In the case of the H$\alpha$ line, the spectrum of the 
$\sigma$ Dra convolved with the corresponding instrumental and rotational kernels was used as reference. For the \ion{Ca}{II} $\lambda$8542, the dashed line indicates the synthetic profile. }
\end{centering}
\end{figure*}

The profiles of the \ion{Na}{I} $\lambda$5889 line from the MRES run are presented in the right column of Fig.~\ref{3} and in the left collumn of Fig.~\ref{5}. 
It is reasonable to suppose that in the spectrum that by December 27, we deal with the pure photospheric profile (plus the possible unresolved interstellar contribution). 
We used this as the reference profile and subtracted it from the other observed profiles. 
The velocities and equivalent widths of the absorption components in the residual spectra were measured with the Gaussian fitting in IRAF \textit{splot} routine.
These measurements are presented in the Table ~\ref{table3}. The EW errors were estimated using the formulae from \citet{Cayrel1988}. Errors were better
than 5\% for the strong lines (EW $\gtrsim$ 100 m\AA) and about 10-15\% for weaker lines.

The first two exposures obtained on December 13 and 15 showed the deep blueshifted profiles, which were the unresolved blends 
of the photospheric and intense circumstellar components. On the night of December 19, we observed the zero-velocity photospheric profile and
one additional component displaced by about $-62$ km s$^{-1}$. 
It is hard to foresee the following evolution of this component because of the missing four observational nights and rapid variability 
in the \ion{Na}{I} D lines. In the spectrum obtained on December 24 one can see a complicated picture in which the photospheric profile appeared with the blue wing distorted 
by the unresolved absorption and another separate component was displaced at  $-108$ km s$^{-1}$. On the successive night the low-velocity ($-36$ km s$^{-1}$) component grew significantly 
and almost reached the saturation. Another component stayed almost at the same velocity, although it also became deeper. 
After one night, on December 27 the intense low-velocity component completely disappeared but the second absorption only slightly reduced in intensity 
and shifted  a few km s$^{-1}$ to the blue. The mean FWHM of this component for three dates (December 24, 25, and 27) was about $25.5$ km s$^{-1}$. 
There is the strong reason to suggest that, despite the missing the night of December 26, we are dealing with a 
slowly evolved high-velocity component forming in the same portion of the gas. On the other hand the low-velocity component showed sufficiently faster evolution on the timescale of about one day.


\begin{figure}
\includegraphics[width=\linewidth, angle =0]{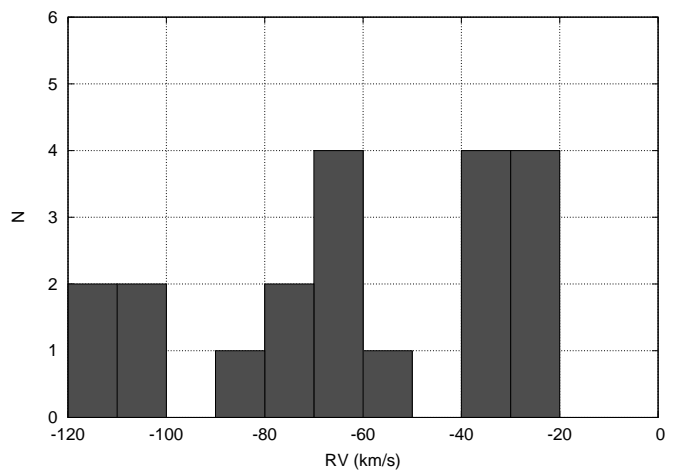}
\caption{\label{4} Distribution of velocities of the additional \ion{Na}{I} D components in the spectra observed with FIES, ESPL, and MRES. }
\end{figure}


The velocity distribution of the \ion{Na}{I} D BACs is shown in Fig.~\ref{4}. From the histogram one can see 
that there are three intervals of radial velocities in which these components appear most frequently. Possibly this means
that the absorption components are formed in different coils of the same gaseous stream with the relatively stable parameters.
Hereafter, based on Fig.~\ref{3} and this histogram, we define the intense and rapidly evolving BACs, from the right 
bin, as low-velocity BACs. We call other components, from the middle and left bins, high-velocity BACs.\\


\begin{table*}
\centering

\caption{Structure of the alkali metals BACs in RZ Psc spectra. The label "a" corresponds to low-velocity BACs, "b" corresponds to high-velocity BACs.
Since there is no possibility of determining the true continuum for the shell component of \ion{Ca}{II} $\lambda$8542,
there are no equivalent width measurements for this line.}
\label{table3}
\begin{adjustbox}{max width=\textwidth}
\begin{tabular}{l  l c c c  c c c  c c }
\hline\hline
\addlinespace

Date&Line&Component&RV (km s$^{-1}$)& W (m\AA) &  Line &RV (km s$^{-1}$)& W (m\AA) &  Line &RV (km s$^{-1}$) \\
\hline
\addlinespace

\addlinespace

2014 Dec 13 & D$_{2}$~ $\lambda$5889 & a & --26  & 343 & \ion{K}{I} $\lambda$7699 & --13 & 88 & \ion{Ca}{II} $\lambda$8542 & --17  \\
                        & & b & --68  & 202 &  & -- & -- & & -- \\
\hline
\addlinespace

\addlinespace
2014 Dec 15 & D$_{2}$~ $\lambda$5889 & a & --25  & 384 & \ion{K}{I} $\lambda$7699 & --3 & 114 & \ion{Ca}{II} $\lambda$8542 & --13 \\
                        & & b & --78  & 133 &  & -- & -- & & -- \\
\hline\addlinespace

\addlinespace
2014 Dec 19 & D$_{2}$~ $\lambda$5889 & a & --  & 17 & \ion{K}{I} $\lambda$7699 & -- & -- & \ion{Ca}{II} $\lambda$8542 & -- \\
                        & & b & --62  & 169 &  & -- & -- & & -- \\
\hline\addlinespace

\addlinespace
2014 Dec 24 & D$_{2}$~ $\lambda$5889 & a & --35  & 383 & \ion{K}{I} $\lambda$7699 & -- & -- & \ion{Ca}{II} $\lambda$8542 & --26 \\
                        & & b & --108 & 299 &  & -- & -- & & -- \\
                        
\hline\addlinespace

\addlinespace
2014 Dec 25 & D$_{2}$~ $\lambda$5889 & a & --36  & 569 & \ion{K}{I} $\lambda$7699 & --40 & 125 & \ion{Ca}{II} $\lambda$8542 & --27 \\
                        & & b & --110  & 340 &  & --112 & 46 & & -- \\
                        

\addlinespace
\hline\addlinespace
2014 Dec 27 & D$_{2}$~ $\lambda$5889 & a & --  & -- & \ion{K}{I} $\lambda$7699 & -- & -- & \ion{Ca}{II} $\lambda$8542 & -- \\
                        & & b & --113  & 277 &  & -- & -- & & -- \\
                        
\hline

\hline
\end{tabular}
\end{adjustbox}
\end{table*}

\subsection{\ion{K}{I} 7699 \AA\ line}

The resonance doublet of the neutral potassium at 7665 and 7699 \AA\ has low ionization potential (4.3 eV) and forms under similar conditions as sodium D lines. Thereby we expect that variability in the potassium lines should resemble the observed variations in \ion{Na}{I} D lines, but scaled according to lower
potassium abundance. The line of \ion{K}{I} at 7665 \AA\ is severely blended by the telluric \element[][][][2]{O}  absorption, while the \ion{K}{I} $\lambda$7699 component 
is more suitable for investigation. Nevertheless, the telluric correction in this spectral range was also performed.

One can see in the third column of Fig.~\ref{5} that the first four spectra demonstrated the single profiles but their depth was variable. Subtraction of the
reference profile  (Dec. 27), as in the case of sodium lines, allowed us to obtain the residual low-velocity circumstellar component observed at December 13,
15, and 25. The values of its RV and EW are also presented in the Table ~\ref{table3}.
On the nights of December 13 and 15 these components appeared slightly shifted shortward as the \ion{Na}{I} D lines. In the spectrum obtained on December 25 the 
two additional BACs were clearly observed. At the same time the intensity of shortward \ion{Na}{I} D components also reached their maximum.

\subsection{\ion{Ca}{II} IR triplet}

The lines of the infrared \ion{Ca}{II} triplet at 8498, 8542, and 8662 \AA\ are clearly observed on the our MRES spectra.
Since the lines of the Paschen series completely disappear in the spectra of dwarfs with spectral type later than G0 \citep{Gray2009}, the effects of blending \ion{Ca}{II} triplet by the 
hydrogen lines in the case of RZ Psc are neglible. Nevertheless, we observed the complicated picture in the calcium lines. For a detailed analysis we choose 
 \ion{Ca}{II} 8542 \AA\ , since its position on the echelle order provides the most reliable continuum tracing in the vicinity of the line. 

The variability of the \ion{Ca}{II} 8542 \AA\ line is
shown in the right column of Fig.~\ref{5}. The outer parts of the line wings coincide with the synthetic profile, while the line core shows significant
deviations from it. On December 19 and
27 the observed profiles appear almost symmetrical but with the core filled by emission. On the other dates the BAC become deeper and
are centered on different velocities, from $-17$ km s$^{-1}$ up to $-27$ km s$^{-1}$. The presence of the \ion{Ca}{II} variable BAC correlates well with the low-velocity component in the \ion{Na}{I} 5889 \AA\ line,
which indicates their common origin.
Subtraction of the synthetic profile from the observed profiles reveals the structure of emission core of the \ion{Ca}{II} 8542 \AA\ line. In Fig.~\ref{6} one
can see the narrow peak close to zero velocity, broad (up to $\pm 100$ km s$^{-1}$) wings, and superimposed
absorption component. While the narrow variable peak most likely arises in the stellar chromosphere, the velocities of the emission wings 
are not typical for chromospheric details and possibly can be attributed to accretion flow. The same is true for the H$\alpha$ line (see below).

\subsection{The H$\alpha$ line}

Figure ~\ref{6} also shows the behavior of the emission component of H$\alpha$ line in RZ Psc spectra. To increase the statistics, the data from 
MRES run were complemented by the two FIES spectra, which are convolved with a corresponding instrumental profile for adjustment of the spectral resolution, and one 
spectrum obtained on October 23, 2012 with the MSS spectrograph of the 6 m BTA telescope by I. Yakunin and I. Potravnov (R$\sim$ 13500). The most notable
feature of the Fig.~\ref{6} is the contrast between the variability of the red and blue emission wings. While the blue emission feature at a velocity of 
about $-70$ km s$^{-1}$ is almost constant, the red emission feature varies considerably. This feature completely disappeared on several dates and appeared on another date.
The equivalent width of the mean emission profile was about 0.5 \AA\ . We discuss the possible interpretation of its shape in the last section of 
the article.


\begin{figure*}
\centering
\includegraphics[width=17cm, angle =0]{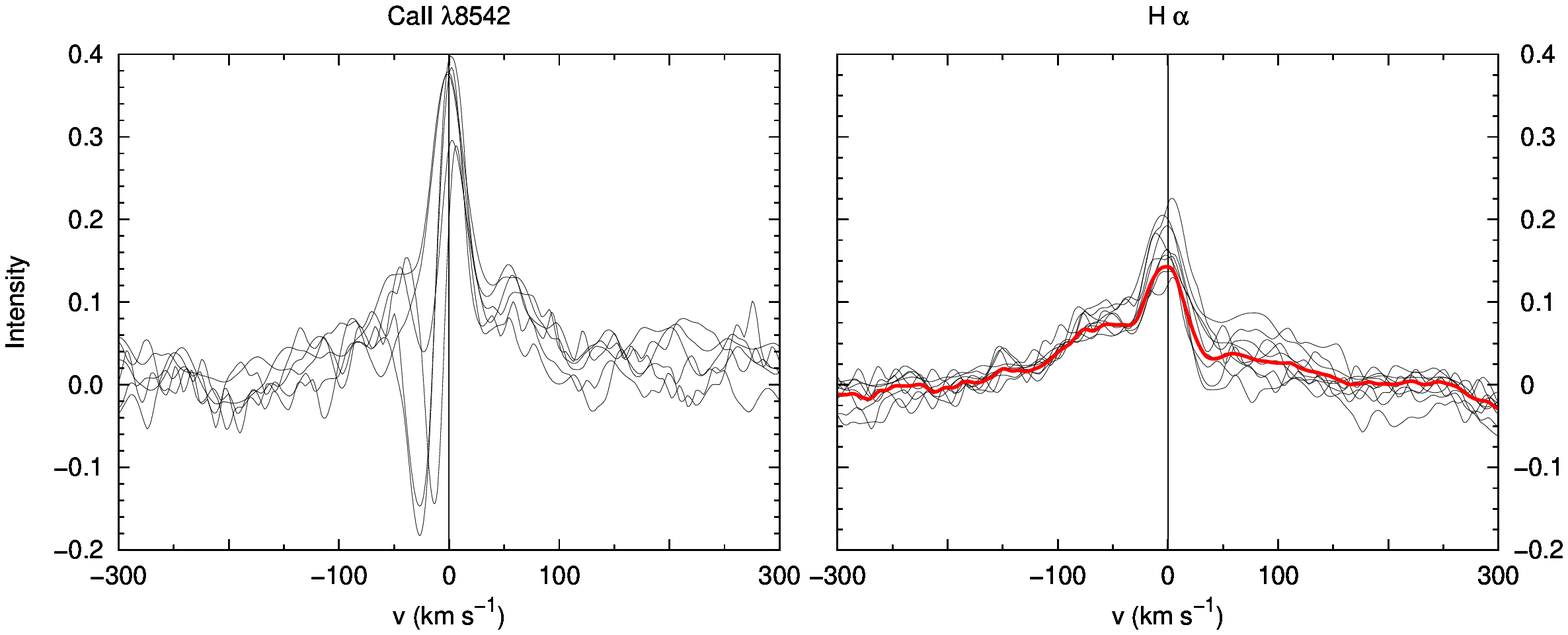}
\caption{\label{6} Residual profiles of  \ion{Ca}{II} 8542 \AA\ and  H$\alpha$ lines. The mean H$\alpha$ emission profile is shown with a thick gray line (red in the electronic version) }
\end{figure*}


\section{Discussion and conclusions}

\begin{figure}
\includegraphics[width=\linewidth, angle =0]{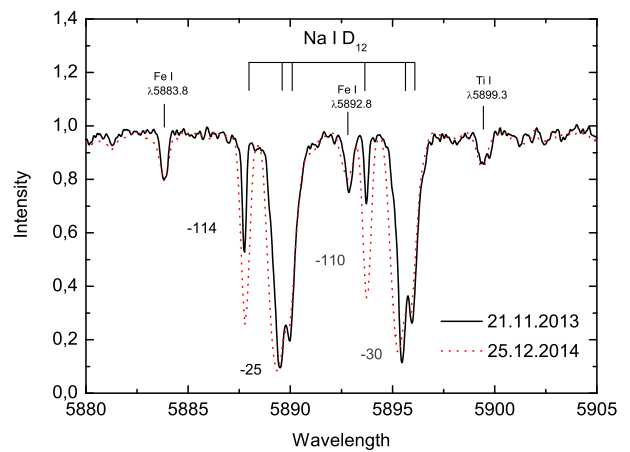}
\caption{\label{7} Comparison of the MRES spectrum obtained on 2014 December 25 (dotted line, red in the electronic version) with the previous spectrum observed on 2013 November 21 
with the FIES spectrograph (solid line). }
\end{figure}

Both of our observational runs (with the ESPL and MRES spectrographs) completely covered the suspected $\sim3.6^{d}$ rotational period of RZ Psc 
(see Section 2). While the first observational set covered successive nights, the second set had some gaps. In both cases we observed the 
prominent evolution of the BACs in the \ion{Na}{I} D lines, but we did not find any clear signs of periodicity in their
appearance. From the other side the comparison of the two spectra from the two different observational sets showed an almost exact coincidence of radial velocities of the BACs observed on December 25, 2014 and November 21, 2013 (Fig.~\ref{7}). 
Apparently, we observed the same gaseous structure where these BACs were formed on these two dates. 
This finding supports our assumption that the variability of the BACs in sodium lines is periodic but
complicated by random fluctuations.

The histogram presented in the Fig.~\ref{4} confirms the issue mentioned above.
From the comparison of the ESPL and MRES 
observations (Fig.~\ref{3}) it is also evident that the star demonstrates the different activity levels during these runs. The ESPL observations showed the location 
of additional components mostly in the intermediate velocity bin in the histogram and the absence of strong absorption
in the velocities of about --20 - 30 km s$^{-1}$. In contrast the MRES spectra demonstrated high-velocity absorption at  $\sim$ --110 km s$^{-1}$ and
a prominent low-velocity component in four of our six spectra.


The theoretical calculations for the magnetospheric accretion model (e.g., \citealt{Muzerolle2001}) show that if the accretion rate is higher than 
$10^{-8}$\Msun yr$^{-1}$ the typical form of the \ion{Na}{I} D profiles is an assymetrical single profile or an inverse P Cyg profile with a blueshifted emission component. The 
emission almost completely disappears in the case of lower accretion rates. But the signs of the matter infall are still preserved. In the case of RZ Psc
we deal with a completely inverse picture. In all of our spectra we observed only signatures of the matter outflow 
and never observed any traces of infall or emission in the lines of the sodium doublet. The absence of the \ion{Na}{I} D
emission can be explained by the low accretion rate (see below). In the framework of the model, which includes both the magnetospheric 
accretion and magnetospheric (or disk) wind, it is possible to explain the formation of the P Cyg-type profile. However, none of the 
existing wind models can explain the appearance of the discrete blueshifted absorption components.
In our previous paper \citep{Grinin2015} we explained the existence of such components as the result of the interaction of 
the circumstellar gas with the inclined magnetosphere in the magnetic propeller regime. The low mass-accretion rate is favorable to realize this mechanism. It arises when the angular velocity of the star (and the magnetosphere) exceeds the angular velocity of the Keplerian 
disk at the truncation radius ($R_{trunk}$). Under this condition, spiral structures in the outflowing matter are formed \citep{Romanova2009}
and most of the accreting gas is scattered into the surrounding space. This issue can explain the total absence of the spectroscopic signatures
of matter infall in \ion{Na}{I} D lines.

The careful consideration of the H$\alpha$ and the IR \ion{Ca}{II} lines in the RZ Psc spectra (Sections 4.4, 4.5) shows that the residual emission in these lines cannot be completely explained by the chromospheric activity. While the narrow peak can be attributed to the chromospheric emission, the broad emission wings (up to $\pm$ 200 km s$^{-1}$) possible arise due to the 
microflares \citep{Montes1998} or accretion \citep{Murphy2011}. 

The knowledge of the stellar distance and radius allows us to estimate the upper limit of the mass accretion rate on the star, using the empirical relationship 
between the observed emission line luminosity and total accretion luminosity. The equivalent width of the mean H$\alpha$ emission in RZ Psc spectra is about 0.5 \AA 
(see \citealt{Grinin2015} and Section 4.5). Using the parameters obtained above and the continuum flux from the
corresponding model of the MARCS grid \citep{Gustafsson2008}, we obtain the H$\alpha$ luminosity $L_{H\alpha}/L_{\sun}\sim 3\cdot10^{-5}$. 
The extrapolation of the relationships from \citet{Rigliaco2012} gives the total accretion luminosity for RZ Psc as $\log (L_{acc}/L_{\sun}) = -3.74$ and the mass accretion rate of 
$\dot{M}$ $\sim$ $7 \cdot 10^{-12}$\Msun yr$^{-1}$. This value is the upper limit of the real accretion rate, since 
there is no certainty that all the H$\alpha$ emission in RZ Psc spectrum is caused by accretion (see the end of the Section 4.4).

Using the same formulae as in \citet{Grinin2015} and the updated parameters of the star, i.e., mass, radius, and accretion rate (Table ~\ref{table2}), we recalculated the corotation and truncation radii for RZ Psc.
We obtained $R_{corr}\sim11R_{\star}$
and $R_{trunk}\sim 30 R_{\star}$ (assuming B = 1kGs). 
This inequality ($R_{trunk} > R_{corr}$) persists even if the stellar magnetic field B is about few hundreds gauss, which supports our previous suggestion 
about the origin of outflow from RZ Psc as a result of the action of the propeller effect. 
The low accretion rate in the system is possibly caused by the existence of an inner cavity in the disk \citep{deWit2013}. This cavity was possibly cleared by planet formation as likely occurs in  transitional disks. The instability and fluctuations of the mass accretion rate can disturb the expected periodicity of the BACs variations
expected in our model. Therefore, the observed picture is much more complex and final conclusions can be drawn only after
collecting more extended observational material.

Thus, RZ Psc has the following set of key evolutionary parameters:

\begin{itemize}

\item RZ Psc is  in the late phases of PMS evolution. The star lies still slightly above the MS on the HR diagram (Fig.~\ref{1}) .\\

\item  The previous item is in agreement with our estimations of the lithium and kinematical age of the star \citep{Grinin2010,Potravnov2013kin}
and with the very low accretion rate.\\
 
\item RZ Psc is surrounded by remnants of the protoplanetary disk. The spectroscopic signatures of circumstellar gas are very weak in comparison with the actively accreting
classical T Tauri stars, but exceed the circumstellar activity footprints in more evolved systems with debris disks. The structure and
kinematics of the BACs observed in RZ Psc spectra strongly resemble  such BACs in the spectra of some actively accreting PMS stars  \citep{Grinin2015}. \\

\item  The IR data show that the circumstellar disk around RZ Psc is in a more advanced evolutionary stage, than young protoplanetary disks. It has an inner gap and a large amount of warm dust \citep{deWit2013}.\\
 
\end{itemize}

We deduce that RZ Psc is a high-latitude isolated post-T Tauri and post-UX Ori-type star whose disk is in the transitional evolutionary stage between primordial and debris disks.
The evolutionary status and observational 
characteristics of RZ Psc make this object a point of a great interest in studying the very late phases of accretion and outflow processes
in the circumstellar disk around young stars near the main sequence.

\begin{acknowledgements} 

This work was supported by grant RFBR 15-02-09191 (V. Grinin, I. Potravnov, D. Shakhovskoy) and in part by the RFBR 15-02-05399 (I. Potravnov and D. Shakhovskoy).
V. Grinin and I. Potravnov were also supported by the Programme of Fundamental Research of the Russian Academy of Sciences 7P ``Experimental and theoretical research of the objects of Solar system and planetary systems of stars''.
We are grateful to P.P. Petrov for the useful discussions and to the referee, whose comments have helped to  improve the paper significantly.  
The assistance of S.P. Belan in photometric observations is also gratefully acknowledged.

\end{acknowledgements}

\bibliographystyle{aa.bst}
\bibliography{reference.bib}
\end{document}